\documentclass[runningheads,fleqn]{svmult}
\usepackage{graphicx} 
\usepackage{subeqnar}  
\usepackage{multicol}
\usepackage{physproc}
\usepackage{cite}

\newcommand{\eq}[1]{Eq.(\ref{#1})}
\newcommand{\fig}[1]{Fig.\ref{#1}}
\newcommand{\sect}[1]{Section~\ref{#1}}
\newcommand{\app}[1]{Appendix~\ref{#1}}

\newcommand{\figwide}{90mm}
\newcommand{\figslim}{55mm}

\newcommand{\avg}[1]{\langle #1 \rangle}

\begin{document}

\title{Entropy Driven Phase Separation}

\author{Richard Vink}

\institute{ 
  Institut f\"{u}r Physik \\ 
  Johannes Gutenberg-Universit\"{a}t \\
  Mainz (Germany)}

\maketitle

\begin{abstract}A grand canonical Monte Carlo method for the simulation of
a simple colloid-polymer mixture called the AO model will be described.
The phase separation known to occur in this model is driven by entropy.
The phase diagram of the unmixing transition, the surface tension and the
critical point will be determined. \end{abstract}

\section{Introduction}

Mixtures of particles are all around us. If you pour oil and water
together you create a mixture. Another example is a solution containing
colloids and polymers. Mixtures will sometimes phase-separate or unmix.
When this happens particles of the same kind cluster together. A famous
example is a mixture of oil and water: after unmixing, the lower part of
the container contains mostly water and the upper part mostly oil with 
an interface in between.

There are different reasons for a mixture to unmix. For instance, the unmixed
state might carry a lower energy. As an example consider a mixture of $A$ 
and $B$ particles interacting with the following pair potentials:
\begin{equation}
  u_{AA}(r) = u_{BB}(r) = 0, \hspace{10mm} u_{AB}(r) = \frac{\epsilon}{r},  
\end{equation}
where $r$ is the distance between two particles and $\epsilon$ some
positive constant. In this mixture, $A$ particles do not feel each other,
and neither do $B$ particles, but when $A$ and $B$ particles are close
together there is an energy penalty. At low temperature, the system tries
to minimize the energy and can only do so by moving the $A$ and $B$
particles as far apart as possible. The system thus unmixes.

There is another mechanism that can induce unmixing. This mechanism has
its origin in entropy and not in energy. The unmixing of colloid-polymer
mixtures for example is driven by entropy. To show that this must be the
case, it is instructive to briefly consider the nature of the interactions
in a typical colloid-polymer mixture. To a crude approximation the
colloids behave as hard spheres. The polymer interactions are more
complicated. A real polymer consists of a chain of bonded monomers. In
principle all monomers should be taken into account explicitly. However,
under certain conditions it is reasonable to describe the polymer chain
with only the coordinate of its center of mass and some effective radius
called the radius of gyration. In 1954 Asakura and
Oosawa~\cite{asakura1954a} proposed a simple model for colloid-polymer
mixtures based on precisely these approximations. In this model colloids
and polymers are treated as spheres with respective radii $R_c$ and $R_p$.
Hard sphere interactions are assumed between colloid-colloid (cc) and
colloid-polymer (cp) pairs while polymer-polymer (pp) pairs can
interpenetrate freely. This leads to the following pair potentials:
\begin{eqnarray}\label{eq:ao}
u_{cc}(r) &=& \left\{
    \begin{array}{ll}
    \infty & \mbox{for $r<2R_c$} \\
    0      & \mbox{otherwise,}
    \end{array}
  \right. \quad 
u_{cp}(r) =
  \left\{
    \begin{array}{ll}
    \infty & \mbox{for $r<R_c+R_p$} \\
    0      & \mbox{otherwise,}
    \end{array}
  \right. \\
u_{pp}(r) &=& 0 \nonumber,
\end{eqnarray}
with $r$ again the distance between two particles. The above equations define
what is nowadays called the AO model. Note that in 1976 the same model was
proposed again and independently by Vrij~\cite{vrij1976a}.

\begin{figure}[t]
\begin{center}
\includegraphics[width=\figslim]{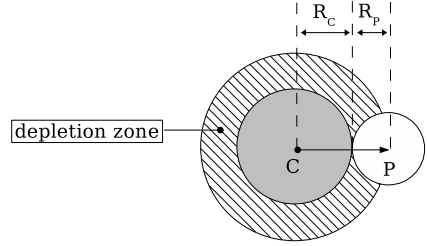}
\caption{~\label{fig:dzone}Colloid and polymer at minimum separation (here and
throughout this text C=colloid and P=polymer). The distance between the centers
of mass equals $R_c+R_p$. This is the minimum allowed separation: any smaller
separation is punished with infinite energy. The colloid thus carries a
spherical depletion zone with volume $V_\delta$ around its center of mass which
cannot contain any polymer centers.}
\end{center}
\end{figure}

\begin{figure}[t]
\begin{center}
\includegraphics[width=\figwide]{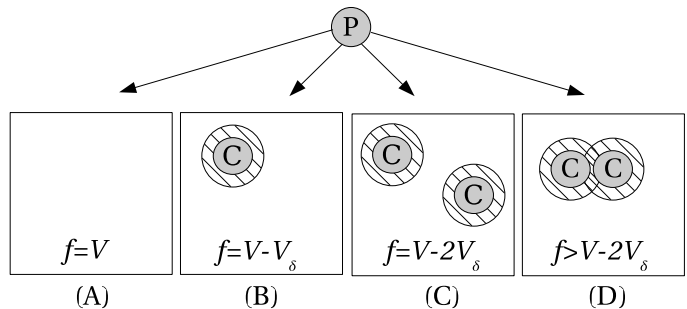}
\caption{~\label{fig:phasesep}Free volume $f$ available to a single
polymer in a container of volume $V$ holding zero, one and two colloidal
particles. See text for details.}
\end{center}
\end{figure}

According to \eq{eq:ao} the energy of a valid AO mixture is always zero.
Therefore, if unmixing occurs (and it does) it cannot be explained in
terms of the energy argument used before. The reason for unmixing is a
little more subtle, see \fig{fig:dzone} and \fig{fig:phasesep}. Shown in
\fig{fig:dzone} is one colloidal particle just touching one polymer. When
the particles touch, the distance between the centers of mass equals
$R_c+R_p$. This is the minimum allowed separation between the particles
because any smaller separation carries an infinite energy penalty. In the
AO model every colloid is thus surrounded by a region which cannot contain
any polymer centers of mass. This region is called the depletion zone. In
three dimensions the volume of the depletion zone equals:
\begin{equation}\label{eq:depl}
  V_\delta = \frac{4\pi}{3} (R_c+R_p)^3.
\end{equation}

Consider now \fig{fig:phasesep} which shows a container of volume $V$
holding zero, one and two colloidal particles. Shown at the top is a
polymer which we want to insert into the container. If the container is
empty we can place the polymer anywhere so the free volume $f$ available
to the polymer is simply $f=V$ (\fig{fig:phasesep}A). If the container
already holds one colloid the free volume decreases to $f=V-V_\delta$
(\fig{fig:phasesep}B). The physics becomes interesting when the container
holds two or more colloids. Two cases can now be distinguished. In the
first case both colloids are well separated and the free volume equals
$f=V-2V_\delta$ (\fig{fig:phasesep}C). In the second case the colloids
are so close together that their depletion zones overlap and the free
volume increases $f>V-2V_\delta$ (\fig{fig:phasesep}D). This immediately
has physical consequences: by clustering together the colloids can
increase the free volume available to the polymers and hence the entropy
of the polymers. Under certain conditions the gain in entropy is
sufficient to drive unmixing.

The AO model thus provides a convenient framework in which to study
entropy driven unmixing. It has sparked much theoretical work and many
simulations~\cite{aarts2002a,roth2000a,schmidt2002a,dijkstra2002a,bolhuis2002a}.  
Despite its apparent simplicity, predicting the phase behavior of the AO
model is no easy task. In this paper an efficient grand canonical Monte
Carlo (MC) method that can be used to simulate the AO model will be
described. The use of the grand canonical ensemble allows one to bypass
certain problems encountered for example in the canonical and Gibbs
ensembles. In particular, one can accurately obtain the surface tension
and also perform simulations close to the critical point. The method will
be used to calculate the phase diagram and the surface tension of the
interface. We will also present an accurate determination of the critical
point using finite-size scaling.

\section{Grand Canonical Monte Carlo}

Imagine an AO mixture contained in some volume $V$ at inverse temperature
$\beta=1/(k_B T)$. In the grand canonical ensemble $V$ and $\beta$ are
fixed but the number of particles inside $V$ is allowed to
fluctuate. The probability of observing a volume containing $N_c$ colloids
and $N_p$ polymers is given by the grand canonical distribution:
\begin{equation}\label{eq:gc}
  P = C z_c^{N_c} z_p^{N_p} e^{-\beta E}, 
\end{equation}
where $C$ is a normalization constant, $E$ the energy given by \eq{eq:ao}
and $\{z_c,z_p\}$ the fugacities of the colloids and polymers,
respectively.

\begin{figure}[t]
\begin{center}
\includegraphics[width=\figslim]{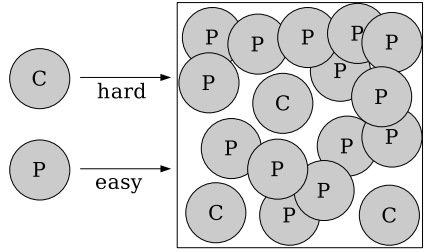}
\caption{~\label{fig:problem}Problem encountered in a standard grand canonical
MC simulation of the AO model. When the number of polymers in the volume is
substantial it becomes difficult to insert additional colloids. Inserting more
polymers remains easy though.}
\end{center}
\end{figure}

In a grand canonical MC simulation of the AO model one would like to
generate configurations of colloids and polymers that sample \eq{eq:gc}.
In the standard approach this is done by attempting to insert a single
particle into $V$ at a random location, or remove a single (randomly
selected) particle from $V$. Insertion and removal are usually attempted
with equal probability and accepted with probabilities that depend on the
energy change of the attempted move and on the fugacities. The standard
approach is however not efficient for the AO model as \fig{fig:problem}
illustrates. The figure shows a volume containing a substantial number of
polymers and some colloids. According to \eq{eq:ao} polymers do not
interact with each other so they are happy to overlap. This is what
generally will happen as the figure shows. Unfortunately, it is nearly
impossible to insert an additional colloid into this system: no matter
where the colloid is placed it will likely overlap with at least some of
the polymers, and colloid-polymer overlaps are forbidden by \eq{eq:ao}. A
standard grand canonical MC simulation of the AO model is thus
characterized by a very low acceptance rate of colloid insertions.

The problem illustrated in \fig{fig:problem} is typical of asymmetric
binary mixtures of which the AO model is an example. The standard grand
canonical MC algorithm does not deal well with such mixtures, essentially
because it moves only one particle at a time. A MC move capable of
removing entire clusters of polymers would be much more efficient. By
using such a cluster move the formation of ``holes'' in the ``sea'' of
polymers is enhanced. If the holes are large enough to contain a colloid,
the acceptance rate of colloid insertions will increase. The MC move used
in this work to simulate the AO model is aimed at doing precisely that.

\section{Cluster Moves}
\label{sec:clmv}

\begin{figure}[t]
\begin{center}
\includegraphics[width=\figwide]{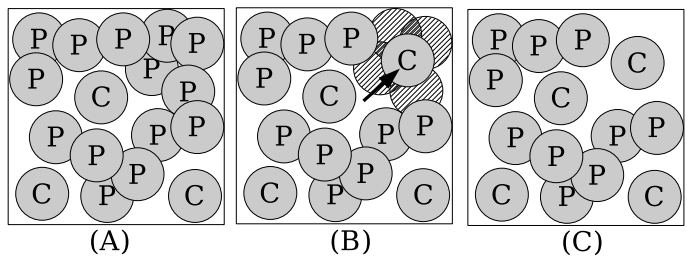}
\caption{~\label{fig:cluster}Inserting an additional colloid into an AO mixture.
(A) The starting configuration. (B) The colloid is inserted at a random location
inside the volume. (C) All polymers that overlap with the colloid are removed.}
\end{center}
\end{figure}

Consider now \fig{fig:cluster}A which again shows an AO mixture of colloids and
polymers. This time we are less shy and simply insert an additional colloid at
some randomly selected location inside $V$. The resulting configuration is shown
in \fig{fig:cluster}B, with the colloid inserted in the upper right corner of
the box. Note that the configuration in \fig{fig:cluster}B is not a valid AO
configuration because the colloid overlaps with four polymers. To fix this the
overlapping polymers are simply removed, leading to the configuration shown in
\fig{fig:cluster}C.

The MC move of \fig{fig:cluster} is of course not sufficient to carry out
a simulation. We also need a reverse MC move capable of bringing us back
from the configuration of \fig{fig:cluster}C to the starting configuration
of \fig{fig:cluster}A. In constructing the reverse move, the key idea is
that for every colloid you take out, a number of polymers must be inserted
into the empty depletion zone left behind by the colloid. But exactly how
many polymers?

To answer this question it is instructive to consider an AO model containing
only polymers. In this case the AO model reduces to a very simple system, namely
that of an ideal gas (remember that the polymers do not interact with each
other). According to basic statistical mechanics, the average density of an
ideal gas is equal to its fugacity. Likewise, for an AO model free of colloids,
the average polymer density equals $z_p$. If we attempt to insert a colloid into
the pure polymer system it will on average overlap with $z_p V_\delta$ polymers
with fluctuations in the average that are Poisson like, i.e.~of order $\sqrt{z_p
V_\delta}$. Because of these fluctuations, the number of polymers $n_p$ that we
must insert for every colloid that we take out cannot simply be a constant.
Instead, $n_p$ must be a random variable drawn from some probability
distribution. One choice that works well (see \sect{sec:ers}) is to draw it
uniformly from the interval $n_p \in [0,m\rangle$ (so including $0$ but
excluding $m$) and $m$ an integer given by:
\begin{equation}\label{eq:m}
  m = 1 + \max \left[1, {\rm int}\left( 
  z_p V_\delta + \alpha \sqrt{z_p V_\delta} \right) \right],
\end{equation}
with $\alpha$ a positive constant of order unity you are free to choose ($\alpha
\approx 2.0$ usually gives good results). Recall that $V_\delta$ is the volume
of the depletion zone given by \eq{eq:depl}. 

\begin{figure}[t]
\begin{center}
\includegraphics[width=\figwide]{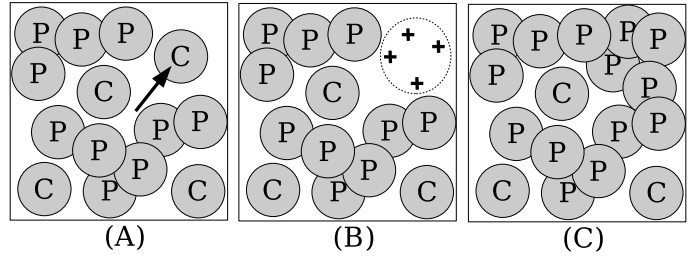}
\caption{~\label{fig:reverse}MC move used to remove a colloid from an AO
mixture. (A) The colloid to be removed is chosen randomly from among those
present. (B) The colloid is removed and $n_p$ random locations are selected
inside the empty depletion zone left behind. The number $n_p$ is uniformly drawn
from the interval $[0,m\rangle$. In this example $n_p=4$. (C) Polymers are
placed onto the random locations.}
\end{center}
\end{figure}

The reverse move can now be constructed and is illustrated in \fig{fig:reverse}.
First, we randomly select a colloid (\fig{fig:reverse}A). The colloid is removed
and $n_p$ random locations inside the depletion zone of this colloid are
selected, with $n_p$ uniformly drawn from the interval $[0,m\rangle$
(\fig{fig:reverse}B). Finally, polymers are placed onto these random locations
resulting in the configuration shown in \fig{fig:reverse}C.

\section{Detailed Balance}

We will now derive the acceptance rates of the MC moves illustrated in
\fig{fig:cluster} and \fig{fig:reverse}. The acceptance rates must be
constructed such that detailed balance is obeyed. This is vital because
the algorithm must sample the grand canonical distribution of \eq{eq:gc}.  
Recall that the condition of detailed balance demands
that~\cite{newman1999a,landau2000a,frenkel2002a}:
\begin{equation}\label{eq:db}
  P(\sigma) g(\sigma \rightarrow \tau) A(\sigma \rightarrow \tau) = 
  P(\tau) g(\tau \rightarrow \sigma) A(\tau \rightarrow \sigma),
\end{equation}
with $P(\sigma)$ the Boltzmann probability of state $\sigma$, $g(\sigma
\rightarrow \tau)$ the probability that the MC scheme generates state
$\tau$ from state $\sigma$, and $A(\sigma \rightarrow \tau)$ the
probability of accepting the new state $\tau$.

The derivation that follows is based on two assumptions:
\begin{enumerate}
\item The integer $m$ of \eq{eq:m} is assumed a constant parameter of the
algorithm. It must be set once at the start of the simulation and it may not be
changed during the course of the simulation. 
\item It is assumed that the insertion of a colloid and the removal of a colloid
are attempted with equal probability.
\end{enumerate}
In an implementation it is important that the above conditions are met. If
they are not the acceptance rates to be derived next may yield wrong
results!

\subsection{Colloid Removal}

The acceptance rate for the removal of a colloidal particle is derived
first. This is the MC move shown in \fig{fig:reverse}. Assume that we
start in a state $\sigma$ containing $N_c$ colloids, $N_p$ polymers and
energy $E_\sigma$. It is convenient to label the state as
$\sigma(N_c,N_p,E_\sigma)$. The energy $E_\sigma$ will of course be zero
but for the derivation it is convenient to write it down explicitly. After
removing the colloid, $n_p$ polymers are inserted into the depletion zone
left behind. The state $\tau$ that we end up in can thus be labeled as
$\tau(N_c-1,N_p+n_p,E_\tau)$. Note that $E_\tau$ need not be zero: if a
second colloid happens to be very close to the colloid that was removed,
it could now overlap with one or more of the $n_p$ polymers that were just
inserted.

To enforce detailed balance the probabilities that appear in \eq{eq:db}
must be written down. We begin with the easy ones $P(\sigma)$ and
$P(\tau)$. These are simply given by the grand canonical distribution of
\eq{eq:gc}:
\begin{equation}
 P(\sigma) = C z_c^{N_c} z_p^{N_p} e^{-\beta E_\sigma}, \hspace{10mm}
 P(\tau) = C z_c^{N_c-1} z_p^{N_p+n_p} e^{-\beta E_\tau}.
\end{equation}

The probabilities $g(\sigma \rightarrow \tau)$ and $g(\tau \rightarrow
\sigma)$ are more complicated. If we are in state $\sigma$ the probability
of ending up in state $\tau$ is given by the product of the probabilities
of the individual steps that were taken in the MC move. After close
inspection of \fig{fig:reverse} the following steps can be identified:
\begin{enumerate}
\item Selecting a colloid at random. Since state $\sigma$ contains $N_c$
colloids the probability of choosing one particular colloid is $1/N_c$.
\item Selecting $n_p$. Since $n_p$ is drawn uniformly from the interval 
$[0,m\rangle$ the probability of this step is $1/m$.
\item Selecting $n_p$ random locations inside a volume $V_\delta$. The
probability of choosing one particular location equals $1/V_\delta$ (see 
\app{sec:rp}). Since $n_p$ locations are selected we must raise this
probability to the power $n_p$. Additionally, we pick up a factorial counting
the number of ways in which the locations can be selected. The total probability
of this step thus becomes $(n_p)!/V_\delta^{n_p}$.
\end{enumerate}
The probability $g(\sigma\rightarrow\tau)$ is thus the product of the 
above three terms.

Finally, we derive $g(\tau\rightarrow\sigma)$. This is the probability
that the reverse MC move brings us back from state $\tau$ to state
$\sigma$. This move is illustrated in \fig{fig:cluster} and upon
inspecting it we observe that state $\sigma$ is regenerated if a colloid
is placed inside $V$ at precisely the same location it was just removed
from. This carries a probability $1/V$ and so we find
$g(\tau\rightarrow\sigma) = 1/V$.

Substitution of the above terms into \eq{eq:db} and using the Metropolis 
choice~\cite{newman1999a} we find that the removal of a colloid must be 
accepted with probability:
\begin{equation}\label{eq:reverse}
  A(N_c \rightarrow N_c-1) = \min \left[1,
  \frac{m N_c}{z_c V}
  \frac{ (z_p V_\delta)^{n_p} }{ (n_p)! }
  e^{-\beta (E_\tau-E_\sigma) } \right].
\end{equation}
Note that moves leading to configurations where $E_\tau$ is not zero are
automatically rejected by the above equation.

\subsection{Colloid Insertion}

To make the algorithm complete the acceptance rate for the insertion of
a colloid must still be derived, i.e.~the MC move of \fig{fig:cluster}. We
can follow the same reasoning as before with perhaps one subtlety. We
start again in a state $\sigma$ containing $N_c$ colloids, $N_p$ polymers
and energy $E_\sigma$, so with label $\sigma(N_c,N_p,E_\sigma)$. All
polymers (say $n_p$ of them) that overlap with the inserted colloid are
removed, so the state $\tau$ that we end up in can be labeled as
$\tau(N_c+1,N_p-n_p,E_\tau)$. Again, the energy $E_\tau$ need not be zero:
if we are unlucky the inserted colloid overlaps with one or more of the
$N_c$ colloids already present. According to \eq{eq:ao} such overlaps also
carry infinite energy.

To construct detailed balance we first write down $P(\sigma)$ and $P(\tau)$:
\begin{equation}
 P(\sigma) = C z_c^{N_c} z_p^{N_p} e^{-\beta E_\sigma},  \hspace{10mm}
 P(\tau) = C z_c^{N_c+1} z_p^{N_p-n_p} e^{-\beta E_\tau}.
\end{equation}

Next, we derive $g(\sigma\rightarrow\tau)$ which is the probability that the MC
move of \fig{fig:cluster} generates state $\tau$ starting in state $\sigma$.
This involves only the random selection of a location inside $V$ so we obtain:
$g(\sigma\rightarrow\tau) = 1/V$.

Finally, we consider $g(\tau\rightarrow\sigma)$. This is the probability that
the reverse MC move of \fig{fig:reverse} transforms state $\tau$ back into state
$\sigma$. This will happen only if the following conditions are met:
\begin{enumerate}
\item The newly inserted colloid is removed again. Since state $\tau$ contains
$N_c+1$ colloids the probability of this step equals $1/(N_c+1)$. 
\item Precisely the same number $n_p$ of polymers that were removed are inserted
again. This step requires some care. Assume that we chose $m$ rather small in
\eq{eq:m}. In that case the forward move of \fig{fig:cluster} might remove more
polymers than the reverse move of \fig{fig:reverse} can possibly put back. This
will happen if $n_p \geq m$. Therefore, the probability $p$ that exactly
the same number of polymers are inserted must be written as:
\begin{equation}
  p = \left\{
    \begin{array}{ll}
      0   & \mbox{if $n_p \geq m$} \\
      1/m & \mbox{otherwise.}
    \end{array}
  \right.
\end{equation}
\item Selecting the same $n_p$ coordinates inside $V_\delta$. As was explained
before this probability equals $(n_p)!/V_\delta^{n_p}$.
\end{enumerate}
The probability $g(\tau\rightarrow\sigma)$ is thus the product of the
above three terms.

Substitution into \eq{eq:db} and using the Metropolis
choice~\cite{newman1999a} we find that the insertion of a colloid must be
accepted with probability:
\begin{eqnarray}
  A(N_c \rightarrow N_c+1) &=& \\
    \left\{
    \begin{array}{ll}
      0 & \mbox{if $n_p \geq m$} \\
      \min \left[1,
      \frac{z_c V}{m (N_c+1)}
      \frac{ (n_p)! }{ (z_p V_\delta)^{n_p} }
      e^{-\beta (E_\tau-E_\sigma) } \right] & \mbox{otherwise.}
    \end{array}
  \right. \nonumber
\end{eqnarray}
Again, moves leading to configurations where $E_\tau$ is not zero are
automatically rejected by the above equation.

\section{Ergodicity}

Now that the MC scheme obeys detailed balance we need to check if it is ergodic.
An algorithm is ergodic if every point in phase space can be reached with finite
probability. It is easy to see that the algorithm is ergodic as far as the
colloids are concerned. For every inserted colloid, a random location inside $V$
was selected so the entire volume is correctly sampled. Similarly, polymers also
sample the entire volume but do so indirectly, namely via the removal of
colloids. Whenever a colloid is removed, a number of polymers are inserted in
the depletion zone. Since the colloids sample the entire volume, so then do the
polymers.

\begin{figure}[t]
\begin{center}
\includegraphics[width=\figwide]{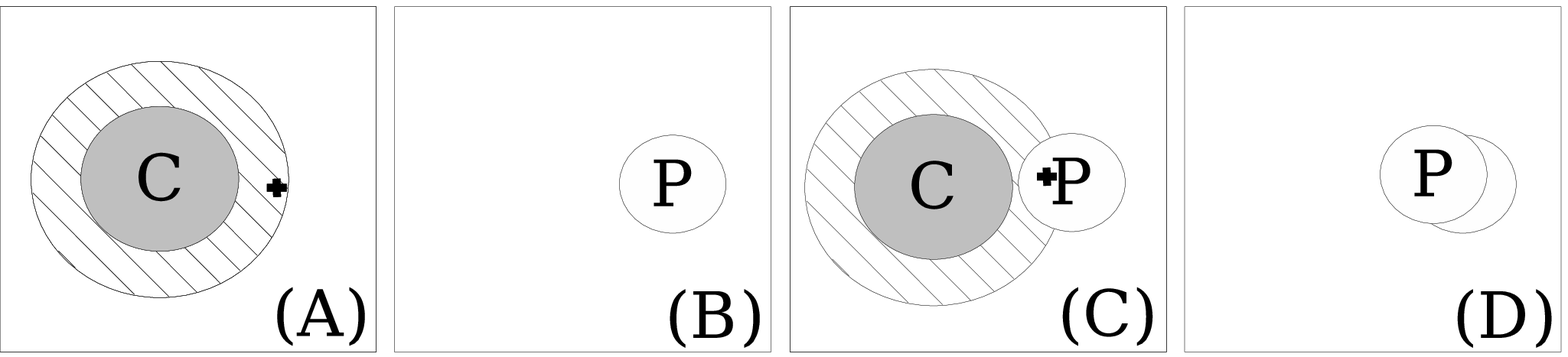}
\caption{~\label{fig:erg}Even low values of $m$ may yield high polymer
densities. (A) The starting configuration showing a colloid with its
depletion zone. The colloid is removed and one polymer is inserted at a
random location inside the depletion zone (cross) resulting in
configuration (B). (C) Another colloid is inserted close to the polymer.
(D) The colloid is removed again with one polymer inserted in its 
depletion zone at the cross resulting in two overlapping polymers.}
\end{center}
\end{figure}

One might argue that the choice of $m$ in \eq{eq:m} violates ergodicity. If at
most $m-1$ polymers are inserted for every colloid, the polymer density will
always be less than $(m-1)/V_\delta$ as a result. While this objection sounds
reasonable it is in fact unjustified as \fig{fig:erg} shows.

\fig{fig:erg} shows an example simulation of the AO model using the cluster
algorithm just described with a low value of $m$, namely $m=2$. This means that
for every removed colloid, at most one polymer is inserted. The starting
configuration is a volume containing one colloid, see \fig{fig:erg}A. The edge
of the depletion zone of the colloid is also drawn. In \fig{fig:erg}B the
colloid is removed and one polymer is placed very close to the edge of the
depletion zone (but still inside of it, of course). The simulation is continued
in \fig{fig:erg}C where another colloid is placed into the system. This colloid
is placed close to the polymer already present, but not too close so the polymer
survives. Finally, in \fig{fig:erg}D the colloid is removed again and one
polymer is placed inside the depletion zone of this colloid, close to the other
polymer. The configuration in \fig{fig:erg}D now shows two polymers practically
stacked on top of each other, even though the algorithm was run with $m=2$. This
removes the above objection.

Running the algorithm with such a low value of $m$ is not recommended though.
The whole point of this discussion is to show that $m$ does not influence the
correctness of the algorithm, only its efficiency.

\section{Early Rejection Scheme}
\label{sec:ers}

In \sect{sec:clmv} it was argued that the number of polymers $n_p$ that one
needs to insert for every colloid that is removed cannot simply be a constant.
In fact, it was shown that $n_p$ is a random variable described by a Poisson
distribution. Yet, in the subsequent description of the algorithm it was decided
to draw $n_p$ uniformly, and not from a Poisson distribution. One might wonder
if the efficiency of the algorithm is not seriously impeded by this choice. The
answer is it is not, provided one implements the so-called early rejection
scheme.

In many MC simulations, the usual approach is to make a change to the
system, calculate the involved energy change, select a random number
between zero and one, compare this random number to the acceptance rate
and finally accept or reject the move. In some cases, particularly in
systems where the interactions are hard sphere like, one can do much
better.

Consider for example the removal of a colloidal particle displayed in
\fig{fig:reverse} and the associated acceptance rate \eq{eq:reverse}. Most
of the quantities appearing in \eq{eq:reverse} are already known at the
start of the move. For example $N_c$, $V_\delta$, $E_\sigma$, $m$, $\beta$
and the fugacities. In fact, the only unknowns are $n_p$ and $E_\tau$. We
also know that if the move is ever going to be accepted $E_\tau$ will be
zero at the end. So the only remaining unknown is $n_p$ which, if you
recall, is a random number drawn from the interval $n_p \in [0,m\rangle$.
The early rejection scheme proceeds as follows:
\begin{enumerate}
\item Select a random number $r$ between zero and one. 
\item Select $n_p$ uniformly from the interval $n_p \in [0,m\rangle$.
\item Calculate the acceptance rate $A(N_c \rightarrow N_c-1)$ using
\eq{eq:reverse} with the above value for $n_p$ and assuming that $E_\tau$ is
zero.
\item If $r>A(N_c \rightarrow N_c-1)$ reject the move immediately otherwise
proceed to the next step.
\item Remove the colloid and insert the $n_p$ polymers. If any of the inserted
polymers produce overlap with other colloids reject the move, otherwise
the move is accepted.
\end{enumerate}
The most CPU time consuming step in the above scheme is step five.
However, this step is only performed for those values of $n_p$ that are
reasonable. Unreasonable values for $n_p$ were already filtered out in
step four at the cost of only a few multiplications and selecting two
random numbers. In this scheme it therefore does not matter so much what
distribution $n_p$ is drawn from. Needless to say, the early rejection
scheme is highly recommended. To speed up the determination of overlap the
link-cell method should also be used.

\section{Application}

In this section the grand canonical cluster scheme will be used to study
bulk phase separation in the AO model. Since the AO energy is either zero
or infinity the temperature plays no role so we simply put $\beta=1$ in
the acceptance rates. The phase behavior of the AO model is thus fixed by
the colloid to polymer size ratio $q=R_p/R_c$ and the fugacities
$\{z_c,z_p\}$. We consider here a size ratio $q=0.8$ and put $R_c=1$ to
set the length scale. The simulations are performed in a box with edges
$L_x \times L_y \times L_z$ and using periodic boundary conditions.
Following convention we define $\eta_p^r \equiv z_p (4\pi/3)R_p^3$ known
as the polymer reservoir packing fraction. Note that $\eta_p^r$ has no
relation to the number of polymers actually in the system. It is just a
different way of expressing the polymer fugacity $z_p$.

In a naive implementation of the scheme one sets the fugacities
$\{z_c,\eta_p^r\}$ and starts the simulation. As the simulation proceeds
colloids and polymers will enter and leave the box. The crucial quantity to
measure is $P(\eta_c)$ defined as the probability of observing a box with
colloid packing fraction $\eta_c \equiv (4\pi/3)R_c^3 N_c/V$. During the
simulation one thus maintains a histogram counting how often a certain colloid
packing fraction has occurred. 

\begin{figure}[t]
\begin{center}
\includegraphics[width=\figwide]{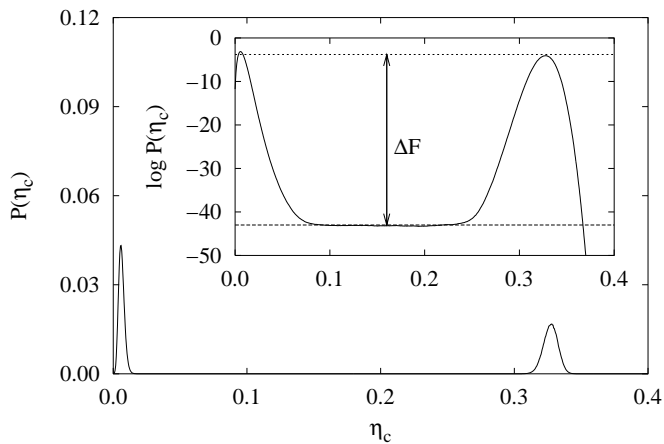}
\caption{~\label{fig:bimodal}Probability $P(\eta_c)$ of observing a
colloid packing fraction $\eta_c$ for an AO mixture with $\{q=0.8;
z_c=87.2; \eta_p^r=1.0\}$ in a simulation of dimensions $L_x=L_y=16.7$ and
$L_z=33.4$. The probability is not normalized. The inset shows the
logarithm of the probability distribution.}
\end{center} 
\end{figure}

If phase separation occurs $P(\eta_c)$ is bimodel. An example distribution is
shown in \fig{fig:bimodal}. The peak at low $\eta_c$ corresponds to the colloid
vapor phase (V), the peak at high $\eta_c$ to the colloid liquid phase (L) and
the region in between is the phase-separated regime. The distribution in
\fig{fig:bimodal} is at coexistence which means that the area under both peaks
is equal. At coexistence, the simulation spends equal time in both phases on
average. Also shown is the logarithm of $P(\eta_c)$. The physical significance
of this curve is its relation to the free energy. The height of the barrier
marked $\Delta F$ corresponds to the free energy barrier separating the phases.
As was shown by Binder~\cite{binder1982a} this barrier is related to the surface
tension via $\gamma = \Delta F/(2A)$ with $A$ the area of the interface.

\begin{figure}[t]
\begin{center}
\includegraphics[width=\figwide]{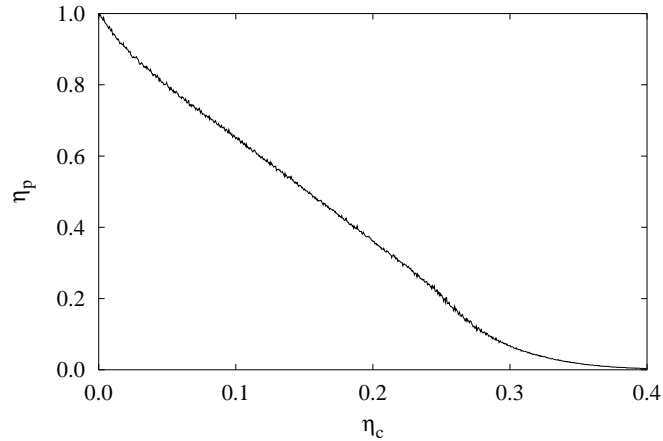} 
\caption{~\label{fig:npavg}Average polymer packing fraction $\eta_p$ as a
function of the colloid packing fraction $\eta_c$ for an AO mixture at
coexistence with $q=0.8$ and $\eta_p^r=1.0$.}
\end{center}
\end{figure}

It is also interesting to measure the average polymer packing fraction $\eta_p
\equiv (4\pi/3)R_p^3 N_p/V$ as a function of $\eta_c$. This result is shown in
\fig{fig:npavg}. In the pure polymer phase ($\eta_c=0$) we expect
$\eta_p=\eta_p^r$ because the polymers then mimic the ideal gas. This is
precisely what \fig{fig:npavg} shows. As the colloid packing fraction increases
the polymer packing fraction in the system decreases to zero.

In the AO model $\eta_p^r$ is the control parameter, much like temperature
is for fluid-vapor transitions. To obtain the phase diagram one simply has
to measure $P(\eta_c)$ at coexistence for a number of different
$\eta_p^r$. The problem in a simulation is finding the value of $z_c$ that
yields coexistence for the chosen $\eta_p^r$ of interest. Additionally, if
the barrier $\Delta F$ is high, it will be difficult to sample $P(\eta_c)$
in the region between the peaks. Fortunately, an array of techniques is
available to overcome these problems~\cite{wilding2001a}. The results in
this work were obtained using a new technique called {\it Successive
Umbrella Sampling}~\cite{virnau2003a}.

\begin{figure}[t]
\begin{center}
\includegraphics[width=\figwide]{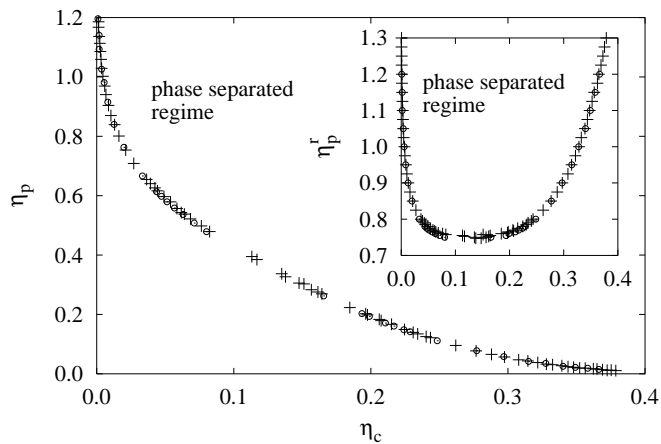} 
\caption{~\label{fig:phase}Phase diagram of the AO model with $q=0.8$ in system
representation. Crosses were obtained using box dimensions $L_x=L_y=16.7$ and
$L_z=33.4$; open circles were obtained in a smaller box with dimensions
$L_x=L_y=13.3$ and $L_z=26.5$. The inset shows the phase diagram in reservoir
representation.}
\end{center}
\end{figure}

\begin{figure}
\begin{center}
\includegraphics[width=\figwide]{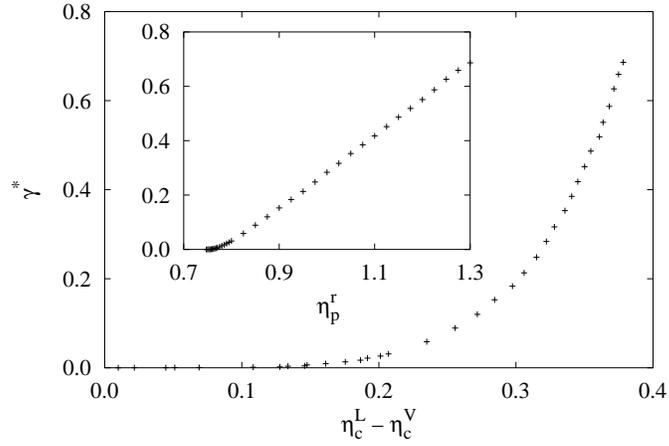}
\caption{~\label{fig:gamma}Reduced surface tension $\gamma^* \equiv 4
R_c^2 \gamma$ for an AO mixture with $q=0.8$ as a function of the
difference in the colloid packing fractions in the coexisting liquid
and vapor phases. The box dimensions were $L_x=L_y=16.7$ and $L_z=33.4$. The
inset shows $\gamma^*$ as a function of $\eta_p^r$.}
\end{center}
\end{figure}

For each $P(\eta_c)$ at coexistence one reads off the colloid packing fraction
of the vapor phase $\eta_c^{\rm V}$ and of the liquid phase $\eta_c^{\rm L}$ and
plots the two points $(\eta_c^{\rm V},\eta_p^r)$ and $(\eta_c^{\rm L},\eta_p^r)$
in a graph. This yields the phase diagram in reservoir representation shown in
the inset of \fig{fig:phase}. By using \fig{fig:npavg} we can convert the
reservoir representation into the experimentally more relevant
$\{\eta_c,\eta_p\}$ or system representation also shown in \fig{fig:phase}. For
every $P(\eta_c)$ one also obtains a value for the surface tension using the
method of Binder. The results of this procedure are shown in \fig{fig:gamma}, in
two different representations.

\begin{figure}
\begin{center}
\includegraphics[width=\figwide]{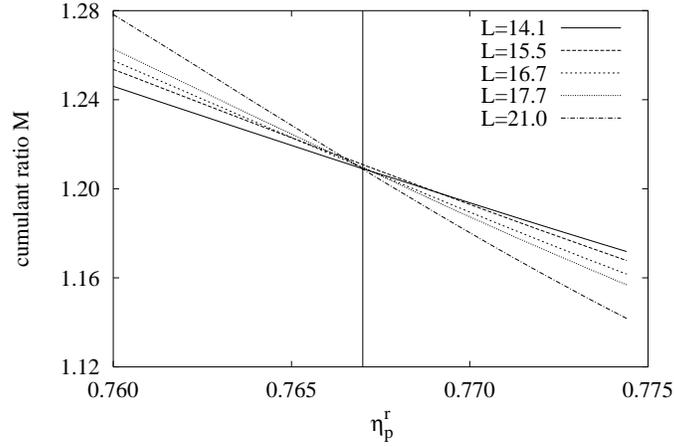}
\caption{~\label{fig:cum}Cumulant ratio $M$ given by \eq{eq:cum} as a function
of $\eta_p^r$ for an AO mixture with $q=0.8$ for various system sizes. The
simulations were performed in a cubic box with edge $L$ as indicated. From the
intercept (vertical line) we obtain for the critical polymer fugacity
$\eta^r_{p,cr}=0.766 \pm 0.002$.} 
\end{center} 
\end{figure}

Finally, we determine the critical polymer fugacity defined as the value
of $\eta_p^r$ above which phase separation begins to take place. From the
inset of \fig{fig:phase} we see that the critical fugacity is around
$\eta_p^r \approx 0.75$. We have performed a finite size scaling
analysis~\cite{wilding1996a} by measuring the cumulant ratio:
\begin{equation}\label{eq:cum}
  M = \frac{ \avg{ \left( \eta_c-\avg{\eta_c} \right)^2} }{ 
  \avg{|\eta_c-\avg{\eta_c}|}^2 },
\end{equation}
as a function of $\eta_p^r$ close to the critical point for different system
sizes. The results are shown in \fig{fig:cum}. The critical fugacity is at the
intersection of the lines from which we obtain $\eta^r_{p,cr}=0.766 \pm 
0.002$.

\section{Conclusions}

The grand canonical cluster scheme described in this chapter is very
successful at modeling phase separation in AO mixtures. In fact, at the
time of writing, it is unsurpassed in speed and accuracy by other
simulation methods~\cite{vink2003a}. However, the method does have its
limitations. If the packing fraction of the colloids is high (say 0.40 and
above) the algorithm is no longer efficient. In that case the insertion of
colloidal particles fails, not because of overlap with polymers, but
because of overlap with other colloids. This problem is well known in hard
sphere simulations.

The algorithm could be improved for systems where the polymer-polymer
interaction is not zero. In these cases the random insertion of $n_p$
polymers into the depletion zone of a colloid may no longer be efficient.
Fortunately, the algorithm is easily adapted to include smarter insertion
moves such as configurational bias~\cite{siepmann1992a}, recoil
growth~\cite{consta1999a}, wormhole MC~\cite{houdayer2002a} and perhaps
PERM~\cite{hsu2003a}. Work along these lines is in progress.

Note also that the presented method is general. The derivation of detailed
balance for example can easily be modified to an AO model confined between
walls or to systems interacting with smooth potentials. This would be the 
subject of further work.

\subsubsection{Acknowledgments} 

I am grateful to the Deutsche Forschungsgemeinschaft for support (TR6/A5)
and to J. Horbach, K. Binder, M. M\"{u}ller, M. Schmidt and P.  Virnau for
collaboration and discussion.

\section{Appendix: Random Points}
\label{sec:rp}

Here we argue that the selection of a random point inside a volume $V$
carries a probability $1/V$. The question is most easily answered by
assuming that the volume is made up of many tiny cells, each with a volume
$\Gamma$. The total number of cells in the volume is thus equal to
$n=V/\Gamma$ and the probability of picking one of them is $1/n=\Gamma/V$.

The difficulty lies in choosing $\Gamma$. In statistical mechanics it is
assumed that the smallest element of phase space has a volume determined
by the thermal wavelength $\Lambda$ leading to $\Gamma=\Lambda^3$. This
explains why the acceptance rates of grand canonical MC schemes often
contain the thermal wavelength~\cite{landau2000a,frenkel2002a}.
Unfortunately, in a MC simulation this is not very useful. Since the
thermal wavelength depends on temperature, particle mass and even the
Planck constant, this choice suggests that a simulation of for example
hard spheres is temperature and mass dependent. 

The key observation is that the physics of the system is indifferent to
the choice of $\Gamma$. To see this consider \eq{eq:reverse} which is the
acceptance rate for the removal of a colloidal particle. Strictly
speaking, the volumes $V$ and $V_\delta$ appearing in \eq{eq:reverse} must
be replaced by $V/\Gamma$ and $V_\delta/\Gamma$, respectively. These
additional factors of $\Gamma$ are, however, readily absorbed into the
fugacities $z_c$ and $z_p$. The fugacity is given by $z=\exp(\beta\mu)$
with $\mu$ the chemical potential, so a rescaling of the fugacity merely
shifts $\beta\mu$ by a constant. This constant has no physical consequence
because it in turn shifts the Hamiltonian by a constant. Thus, $\Gamma$
has no physical consequence and may therefore be set to unity which was
done throughout this text.

\bibliographystyle{prsty}

\end{document}